\documentclass{ws-ijmpa}
\usepackage[super,compress]{cite}
\usepackage{graphicx}
\usepackage{hyperref}
\usepackage{lineno}
\usepackage{amsmath}
\begin{document}
\markboth{Barnab\'as P\'orfy for the NA61/SHINE Collaboration}{Femtoscopy at NA61/SHINE using symmetric L\'evy sources
}

%
\catchline{}{}{}{}{}
%

\title{Femtoscopy at NA61/SHINE using symmetric Lévy sources\\ in central $^{40}$Ar+$^{45}$Sc from 40$A$ GeV/\textit{c} to 150$A$ GeV/\textit{c} 
}

\author{Barnab\'as P\'orfy for the NA61/SHINE Collaboration}

\address{HUN-REN Wigner Research Centre for Physics, \\Konkoly-Thege Mikl\'os \'ut 29-33, H-1121 Budapest, Hungary\\
porfy.barnabas@wigner.hun-ren.hu}

\address{Department of Atomic Physics, Faculty of Science, E\"otv\"os Lor\'and University, \\P\'azm\'any P\'eter s\'et\'any 1/A, H-1111 Budapest, Hungary\\
barnabas.porfy@cern.ch}

\maketitle

\begin{history}
\received{Day Month Year}
\revised{Day Month Year}
\end{history}

\begin{abstract}
In the recent decades of high energy physics research, it was demonstrated that strongly interacting quark-gluon plasma (sQGP) is created in ultra-relativistic nucleus-nucleus collisions. Investigation and understanding of properties of the hadronic matter is among the important goals of NA61/SHINE collaboration at CERN SPS. Mapping of the phase diagram is achieved by varying the collision energy (5 GeV $<\sqrt{s_{\textnormal{NN}}}<$17 GeV) and by changing the collision system (p+p, p+Pb, Be+Be, Ar+Sc, Xe+La, Pb+Pb). 

We report on the measurement of femtoscopic correlations in intermediate systems at intermediate SPS energies. Interpreting the results of measurements within the symmetric L\'evy source formalism, we discuss the values of L\'evy source parameters as a function of average pair transverse mass. One of the physical parameters is particularly important, the L\'evy exponent $\alpha$, which describes the shape of the source and may be related to the critical exponent $\eta$ in the proximity of the critical point. Therefore, measuring it may shed light on the location of the critical endpoint of the QCD phase diagram.

\end{abstract}

\ccode{PACS numbers: 25.75.-q, 25.75.Ag, 25.75.Gz,  25.75.Nq}


\section{Introduction} 
\begin{figure}[h!]
\centering
\includegraphics[width=1\textwidth]{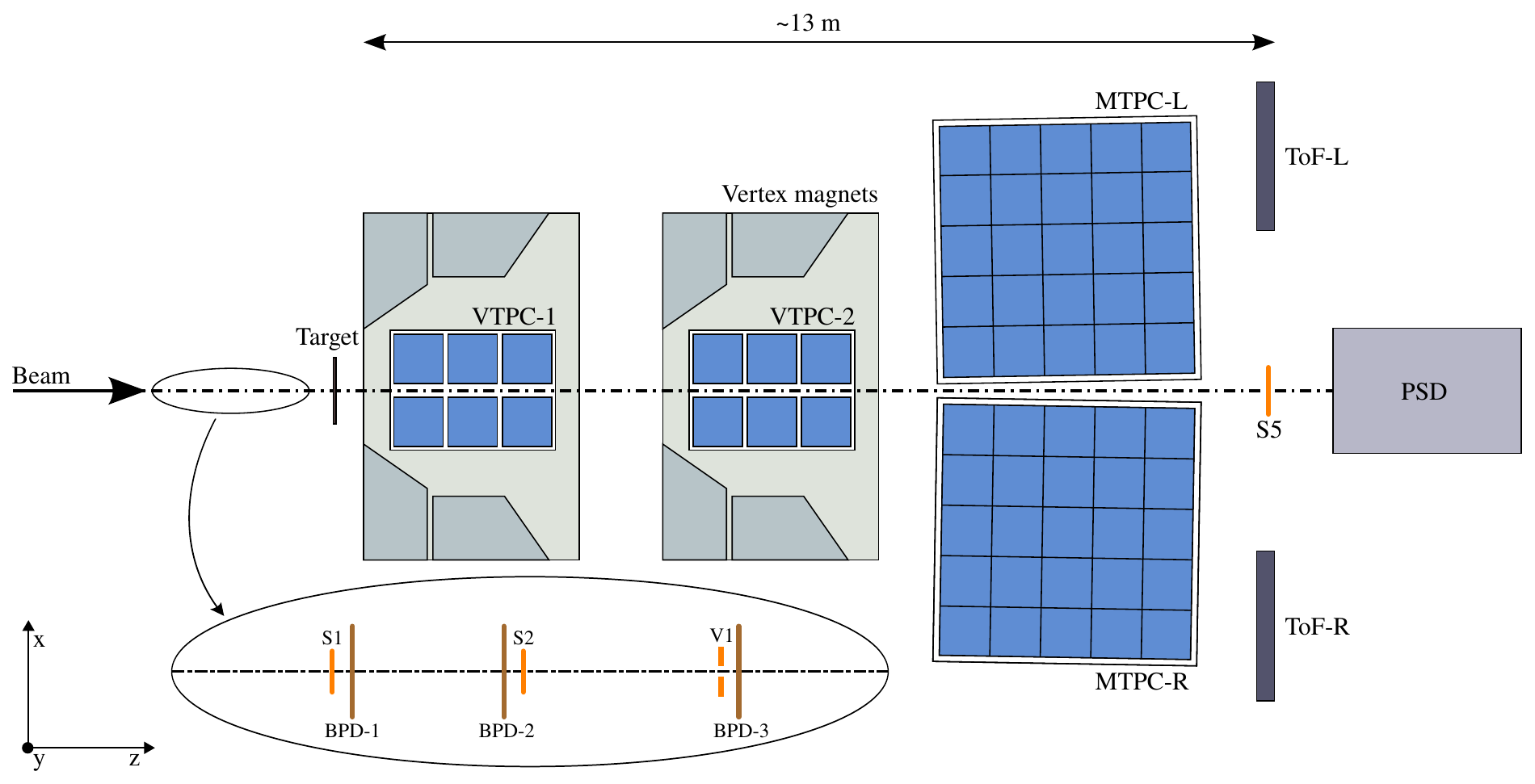}
\caption{The setup of the NA61/SHINE detector system during the run of Ar+Sc.}
\label{fig:detectorsetup}
\end{figure}

The NA61/SHINE is a fixed-target experiment located in the North Area H2 beamline of the CERN SPS. The experiment utilizes multiple Time Projection Chambers (TPCs) that cover the entire forward hemisphere, effectively functioning as a large-acceptance hadron spectrometer, for more details see Ref.~\refcite{Abgrall:2014xwa}. The detector setup of NA61/SHINE,  depicted in Fig.~\ref{fig:detectorsetup},  enables exceptional tracking capabilities down to zero transverse momentum.

The primary objectives of NA61/SHINE include investigating and mapping the phase diagram of strongly interacting matter at various temperatures and baryon-chemical potentials, essentially creating a system and beam energy scan.  This proceedings focuses on two-pion femtoscopy study using intermediate ($^{40}$Ar+$^{45}$Sc) collision system at beam momenta of 40\textit{A} GeV/\textit{c} ($\sqrt{s_{\rm{NN}}} \approx 8.8$ GeV) and 75\textit{A} GeV/\textit{c} ($\sqrt{s_{\rm{NN}}} \approx 12$ GeV), at 0\--10\% centrality, compared to our previous analysis at 150\textit{A} GeV/\textit{c} ($\sqrt{s_{\rm{NN}}} \approx 17$ GeV) in both $^{7}$Be+$^{9}$Be and $^{40}$Ar+$^{45}$Sc.  Our analysis studies the QCD phase diagram using quantum-statistical (Bose-Einstein) correlations with final state interactions, described with spherically symmetric L\'evy distributions~\citen{Csorgo:2003uv,Metzler:1999zz}:
\begin{equation}\label{eq:levydistr}
\mathcal{L}(\alpha,R,\boldsymbol{r})=\frac{1}{(2\pi)^3} \int \rm{d}^3 \boldsymbol{ \zeta} e^{i \boldsymbol{ \zeta} \boldsymbol{r}} e^{-\frac{1}{2}| \boldsymbol{ \zeta} R|^{\alpha}},
\end{equation}
where $R$ is the L\'evy scale parameter, $\alpha$ is the L\'evy stability index, $\boldsymbol{r}$ is the vector of spatial coordinates and $\boldsymbol{ \zeta}$ is the vector that represents the integration variable, all of the equations use $\hbar = 1$.  

The key relationship for measuring Bose-Einstein correlations shows that the momentum correlations, $C(q)$, are related to the properties of the particle-emitting source, $S(x)$, which describes the probability density of particle creation for a relative coordinate \textit{x}. This is expressed as 
\begin{equation}
C(q) \cong 1 + | \tilde{S}(q) |^2,
\end{equation}
where $\tilde{S}(q)$ is the Fourier transform of $S(x)$, and \textit{q} is the relative momentum of the particle pair (with the dependence on the average momentum, $K$, of the pair suppressed. See more details in Ref. ~\citen{NA61SHINE:2023qzr}). 

The L\'evy assumption results in a statistically acceptable description of the source. Two special cases of the distribution in Eq{.}(\ref{eq:levydistr}) are the Gaussian distribution for $\alpha = 2$ and Cauchy distribution with $\alpha = 1$. Unlike the Gaussian assumption, in our case a more general approach is taken by not restricting our assumption to $\alpha = 2$. Non-Gaussianity may arise due to critical fluctuations and the emergence of large-scale spatial correlations~\cite{Csorgo:2008ayr}.  This dependence can be described with L\'evy distribution that shows a power-law tail $\sim r^{-(1 + \alpha)}$ in three dimensions for $\alpha < 2$, where $r \equiv |\boldsymbol{r}|$. Critical behavior is characterized, among other things, by the critical exponent related to spatial correlations, $\eta$, which also follows a power-law tail $\sim r^{(-(d - 2 + \eta)}$, with \textit{d} denoting dimensions. It is suggested in Refs. ~\citen{Halasz:1998qr,Stephanov:1998dy} that the universality class of quantum chromodynamics (QCD) matches that of the 3D Ising model. In the 3D Ising model, $\eta = 0.03631 \pm 0.00003$~\cite{El-Showk:2014dwa}. Alternatively, the random external field 3D Ising model's universality class may be considered, corresponding to $\eta = 0.50 \pm 0.05$~\cite{Rieger:1995aa}. The L\'evy exponent $\alpha$ has been proposed to be directly related to or explicitly equal to the critical exponent $\eta$~\cite{Csorgo:2005it}, in absence of other phenomena affecting the source shape. However, the L\'evy-shape of the source can also be influenced to several different factors besides critical phenomena, including QCD jets, anomalous diffusion, and others~\citen{Metzler:1999zz,Csorgo:2003uv, Csorgo:2004sr, Kincses:2022eqq, Korodi:2022ohn}.  Therefore, while a non-monotonic behavior of $\alpha$ is expected from the above simple picture,  to understand the physical processes influencing the spatial structure of the hadron emission, measurements of $\alpha$ in different collision systems at different energies are needed.  For an overview of recent experimental results, see Ref.~\refcite{Csanad:2024hva}.

The L\'evy exponent of the source distribution can be measured using femtoscopic correlation functions. These correlation functions admit the following form for L\'evy sources:
\begin{equation}\label{eq:corrfunc}
C^0_2(q) = 1 + \lambda \cdot e^{-(qR)^\alpha},
\end{equation}
where $C^0_2(q)$ is the correlation function in absence of interaction and the two physical parameters: $\lambda$ as the correlation strength and $R$ as the HBT scale parameter. The correlation strength $\lambda$ (often called the intercept parameter) is interpreted in terms of the core-halo model in Refs. ~\citen{Csorgo:1999sj,Csorgo:1995bi}. 

At vanishing relative momentum, the correlation function of Eq{.}(\ref{eq:corrfunc}) reaches the intercept value $1 + \lambda$.  Since zero relative momentum is not accessible in the measurements, extrapolation from the region where two tracks are experimentally resolved, is needed. It is commonly observed that the intercept parameter $\lambda$ is less than 1, as also expected from Eq{.}(\ref{eq:lambda}).  It is important to highlight that $\lambda \neq 1$ can also be attributed to other factors, such as coherent pion production~\cite{Csorgo:1999sj,Weiner:1999th} or background noise from improperly reconstructed particles. 

The core-halo model proposes that the source \textit{S} is made up of two parts, the core and the halo $S_{\rm{core}}$ and $S_{\rm{halo}}$, respectively~\cite{Csorgo:1995bi,Csorgo:1999sj}. The core contains pions created close to the center, either directly from hadronic freeze-out (primordial pions) or from extremely short-lived (strongly decaying) resonances. The halo consists of pions created from long-lived resonances and the general background.  It may extend to thousands of femtometers, while core part has a size of around a few femtometers.
The intercept parameter then can be expressed as 
\begin{equation}\label{eq:lambda}
\lambda = \left(\frac{N_{\rm{core}}}{N_{\textnormal{total}}}\right)^2,
\end{equation}
where \textit{N} denotes their respective multiplicities in the given event class and $N_{\textnormal{total}} = N_{\rm{core}} + N_{\rm{halo}}$ contains all pions created.

\section{Measurement details}
We discuss the measurements of one-dimensional two-pion femtoscopy correlation functions for identified pion pairs in Ar+Sc collisions at 40\textit{A} GeV/\textit{c} and 75\textit{A} GeV/\textit{c} with 0-10\% centrality. We investigate the combination of positive pion pairs ($\pi^+\pi^+$) and negative pion pairs ($\pi^-\pi^-$). Event and track quality cuts were applied before particle identification (PID). The PID was performed by measuring the energy loss in the TPC gas (d$E$/d$x$) and comparing it with the Bethe-Bloch curves, following a similar method as detailed in Ref.~\refcite{NA61SHINE:2023qzr}.  The contribution from track merging were eliminated with momentum based two-track distance cut, mentioned in Ref.~\refcite{Czopowicz:2022nzy}.

The centrality interval was selected by measuring the forward energy with the Projectile Spectator Detector~\cite{Abgrall:2014xwa}. The measured pion pairs were grouped into seven average transverse momentum bins ranging from 0 to 500 MeV/c.

\begin{figure}
\centering
\includegraphics[width=.49\textwidth]{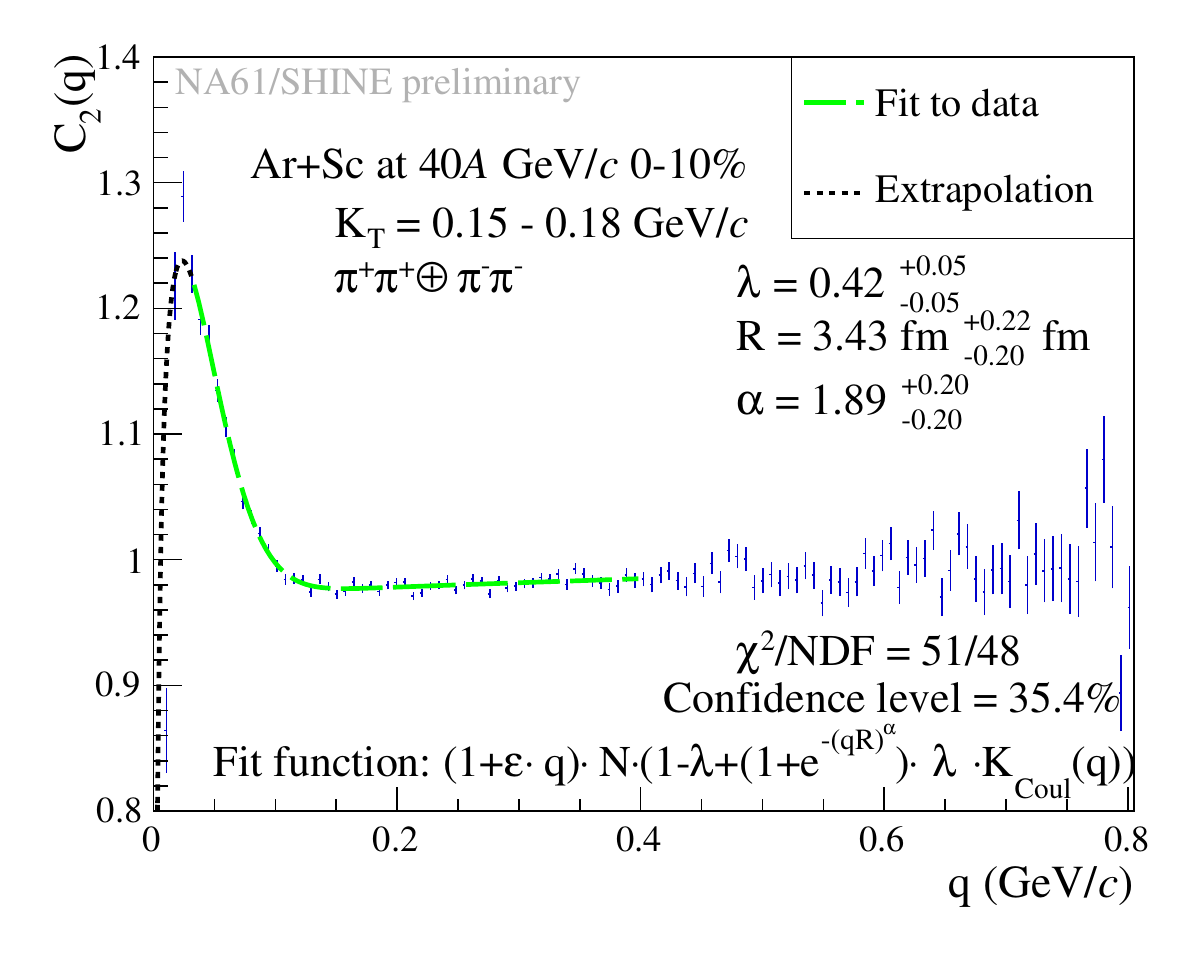}
\includegraphics[width=.49\textwidth]{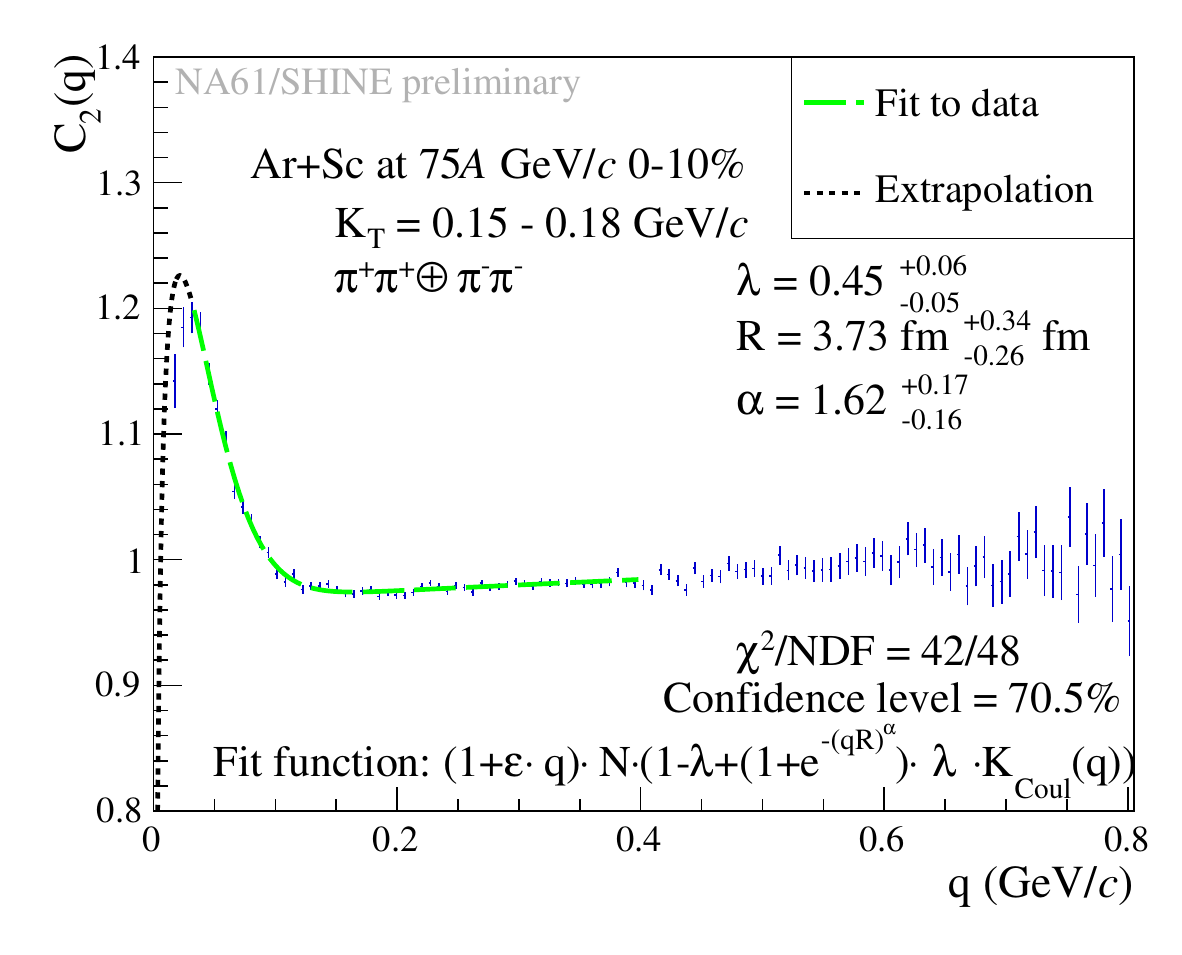}
\caption{Example fit of $\pi^+\pi^+ \oplus \pi^-\pi^-$ femtoscopic correlation function for 40\textit{A} GeV/$c$ and 75\textit{A} GeV/$c$, respectively.  Points with error bars represent the data, the  dashed line shows the fitted function with Coulomb correction. In the low-$q$ region, the dotted line indicates the extrapolated function outside of the fit range.}
\label{fig:examplefit}
\end{figure}

In this analysis,  as mentioned above, we investigate like-charged pion pairs that are influenced by the Coulomb repulsion. The final state Coulomb effect has been neglected in the previously defined correlation function. The correction required by this effect can be done by simply taking the ratio of $C^{\rm{Coulomb}}_2(q)$ and $C^0_2(q)$:
\begin{equation}\label{eq:coulombdef}
K_{\rm{Coulomb}}(q) = \frac{C^{\rm{Coulomb}}_2(q)}{C^0_2(q)},
\end{equation}
where $C^{\rm{Coulomb}}_2(q)$ is the interference of solutions of the two-particle Schr\"odinger equation with a Coulomb-potential~\cite{Kincses:2019rug,Csanad:2019lkp}. The numerator in Eq{.}(\ref{eq:coulombdef}) cannot be calculated analytically and requires numerical calculations. Moreover, the calculations have to account for the assumption of L\'evy-shaped sources. Therefore a novel numerical calculation of the Coulomb final-state interaction with Bose-Einstein correlation functions is used, as presented in Ref. ~\refcite{Nagy:2023zbg}.
Then, one can modify the correlation function, Eq{.}(\ref{eq:corrfunc}), coupling it with the core-halo picture by utilizing the Bowler-Sinyukov method~\cite{Sinyukov:1998fc, Bowler:1991vx}. The halo part contributes at very small values of relative momenta, \textit{q}. Therefore, it does not affect the source radii of the core part~\cite{Maj:2009ue}. Then the correlation function takes the following form:
\begin{equation}\label{eq:fitfunc}
C_2(q) = N\cdot \left( 1 + \epsilon \cdot q\right) \cdot \left(1 - \lambda + \lambda \cdot \left(1 + e^{-|qR|^\alpha} \right) \cdot K_{\rm{Coulomb}}(q) \right),
\end{equation}
where $N$ is introduced as normalization parameter and $K_{\rm{Coulomb}}(q)$ handles the Coulomb interaction, additionally $\varepsilon$ is introduced to describe a possible long-range linear background in the form of $\left(1 + \varepsilon \cdot q \right)$.
It is important to highlight that the Coulomb correction is calculated in the pair-center-of-mass (PCMS) system, while the measurement is often done in the longitudinally co-moving system (LCMS). The assumption of Coulomb correction for the one-dimensional Bose-Einstein correlations in LCMS picture is that the shape of the source is spherical, i.e.  $R_{\rm{out}} = R_{\rm{side}} = R_{\rm{long}} = R \equiv R_{\rm{LCMS}}$.  However the shape of the source, is not spherical in the PCMS. Therefore, an approximate one-dimensional size parameter,$R_{\rm{wave}}$, is needed that can be defined as in Ref.~\refcite{Kurgyis:2020vbz} and expressed with the following formula:
\begin{equation}
\bar{R}_{\rm{PCMS}} = \sqrt{\frac{1 - \frac{2}{3}\beta_{\rm{T}^2}
}{1-\beta^2_{\rm{T}}}} \cdot R,
\end{equation} 
where $\beta_{\rm{T}} =  \frac{K_{\rm{T}}}{m_{\rm{T}}}$.  
Moreover, momentum difference in the Coulomb correction is in PCMS as a function of $q_{\textnormal{PCMS}}=q_{\textnormal{inv}}$ (the invariant four-momentum difference equals the three-momentum difference in the PCMS). 
Furthermore, one would need to perform the measurement as a function of both $q$ and $q_{\textnormal{inv}}$.  Following the estimation in Ref.~\refcite{Kurgyis:2020vbz}, a simple, approximate relation of the two may be given as $q_{\textnormal{inv}} \approx \sqrt{1-\beta_{\textnormal{T}}^2/3}\cdot q$. Implementing both of the above mentioned effects results in the following formula for the Coulomb correction expressed in terms of $q$
and $R_{\textnormal{LCMS}}$, based on the 3D calculation in PCMS:
\begin{equation}\label{e:KCoulombfinal}
    K_{\textnormal{Coulomb}}\left(q,R\right) = 
    K_{\textnormal{Coulomb}}^{3D,\;PCMS}\left(\sqrt{1-\frac{\beta_{\textnormal{T}}^2}{3}}\cdot q, \sqrt{\frac{1-\frac{2}{3}\beta^2_{\textnormal{T}}}{1-\beta^2_{\textnormal{T}}}} \cdot R\right),
\end{equation}
To illustrate, two example fits using Eq{.}(\ref{eq:fitfunc}) are shown on Fig.~\ref{fig:examplefit}.  Based on the goodness-of-fit of the measured correlation functions, the L\'evy-stable source assumption is statistically acceptable.  

\section{Results}
\begin{figure}[b]
\centerline{\includegraphics[width=1.\textwidth]{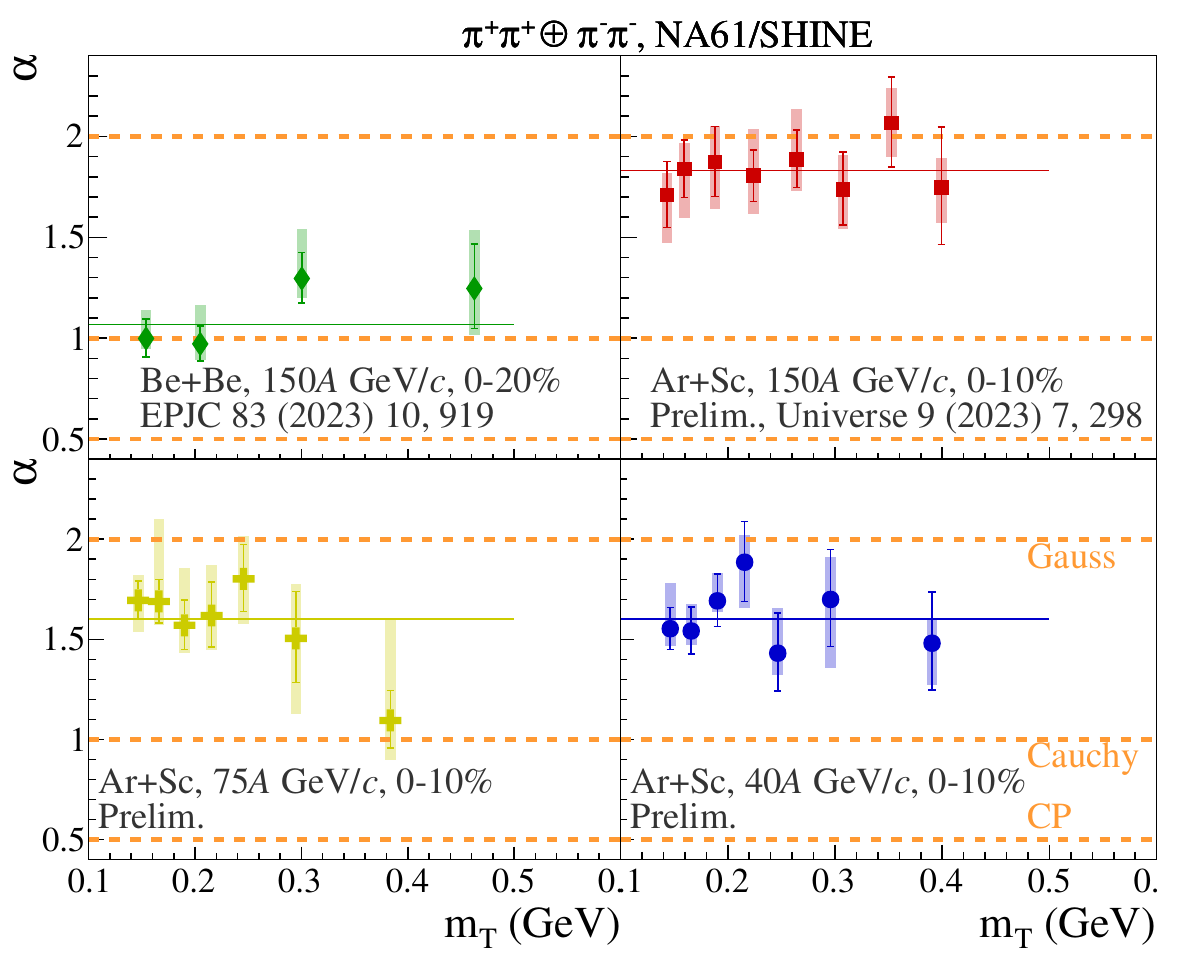}}
\caption{The fit parameters, for $0$--$20$ \% central Be+Be at 150\textit{A} GeV/\textit{c} and $0$--$10$\% central Ar+Sc at 150\textit{A} GeV/\textit{c}, 75\textit{A} GeV/\textit{c}, and 40\textit{A} GeV/\textit{c}, as a function of $m_{\rm{T}}$. Special cases corresponding to a Gaussian ($\alpha=2$) or a Cauchy ($\alpha=1$) source are shown, as well as $\alpha=0.5$, the conjectured value corresponding to the critical endpoint, while the constant $\alpha$ fit is shown with a solid line. Boxes denote systematic uncertainties, bars represent statistical uncertainties.\label{fig:resultsA}}
\end{figure}

\begin{figure}[b]
\centerline{\includegraphics[width=1.\textwidth]{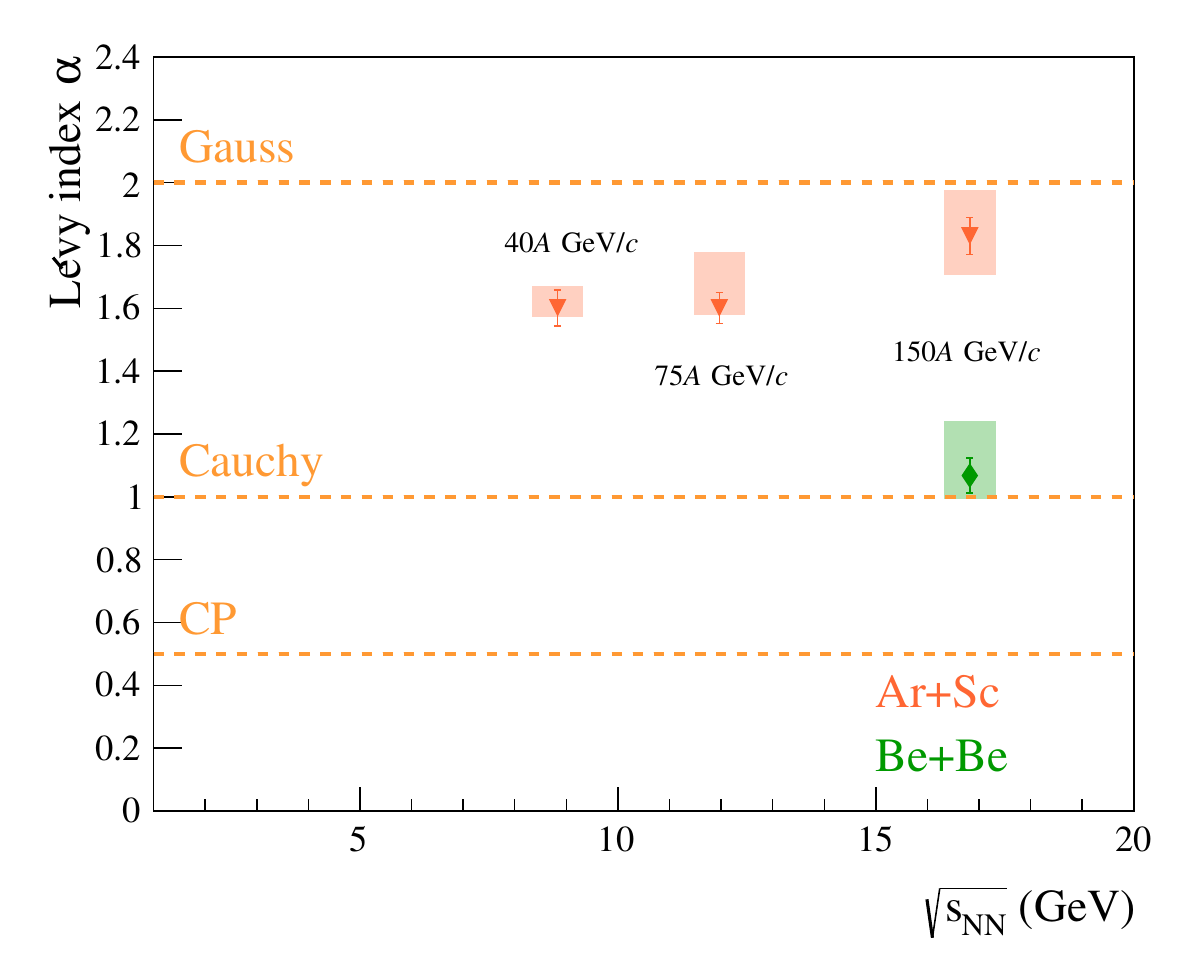}}
\caption{The constant fit to $\alpha$ , for $0$--$20$ \% central Be+Be at 150\textit{A} GeV/\textit{c} and $0$--$10$\% central Ar+Sc at 150\textit{A} GeV/\textit{c}, 75\textit{A} GeV/\textit{c}, and 40\textit{A} GeV/\textit{c}, as a function of $\sqrt{s_{\rm{NN}}}$.  Special cases corresponding to a Gaussian ($\alpha=2$) or a Cauchy ($\alpha=1$) source are shown, as well as $\alpha=0.5$, the conjectured value corresponding to the critical endpoint. Boxes denote systematic uncertainties, bars represent statistical uncertainties.\label{fig:resultsAfit}}
\end{figure}

The three physical parameters ($\alpha$, $R$, and $\lambda$) were measured in seven bins of pair transverse momentum,  $K_{\rm{T}}$. The three mentioned parameters were obtained through fitting the measured correlation functions with the formula shown in Eq{.}(\ref{eq:fitfunc}). The results were investigated regarding their transverse momentum dependence. In the following, we discuss the transverse $q$ mass dependence of $\alpha$, $R$, and $\lambda$; where transverse mass is expressed as $m_{\rm{T}} = \sqrt{m^2c^4+K_{\rm{T}}^2c^2}$ and where $m$ is the particle mass.
As explained above, the shape of the source is often assumed to be Gaussian. The L\'evy stability exponent, $\alpha$, can be used to extract the shape of the tail of the source. Our results for 40\textit{A} GeV/$c$ and 75\textit{A} GeV/$c$, presented in bottom part of Fig. ~\ref{fig:resultsA}, and show $\alpha$ values between 1.5 and 2.0. The measured values are comparable to the highest beam energy result in Ar+Sc at 150\textit{A} GeV/$c$~\cite{Porfy:2023yii}, which imply a source closer to the Gaussian shape ($\alpha = 2.0$ case).  Contrary to the one in Be+Be collisions~\cite{NA61SHINE:2023qzr} which shows approximately Cauchy shaped source ($\alpha = 1.0$ case).  The observed $\alpha$ parameter is also significantly higher than the conjectured value at the CP ($\alpha \approx 0.5$).  Assuming no dependence for the shape of the source $\alpha$ on m$_{\textnormal{T}}$, one can fit a constant to the parameter values. In our case, we performed the fit using only the statistical uncertainties of $\alpha$ parameter values, shown in Fig.~\ref{fig:resultsAfit}.  In the lower energies of Ar+Sc system, an interesting, decreasing trend with decreasing collision energy seem to be developing which will be explored further.  Other experimental results can be found in Ref.~\refcite{Csanad:2024hva}. 

\begin{figure}[b]
\centerline{\includegraphics[width=1.\textwidth]{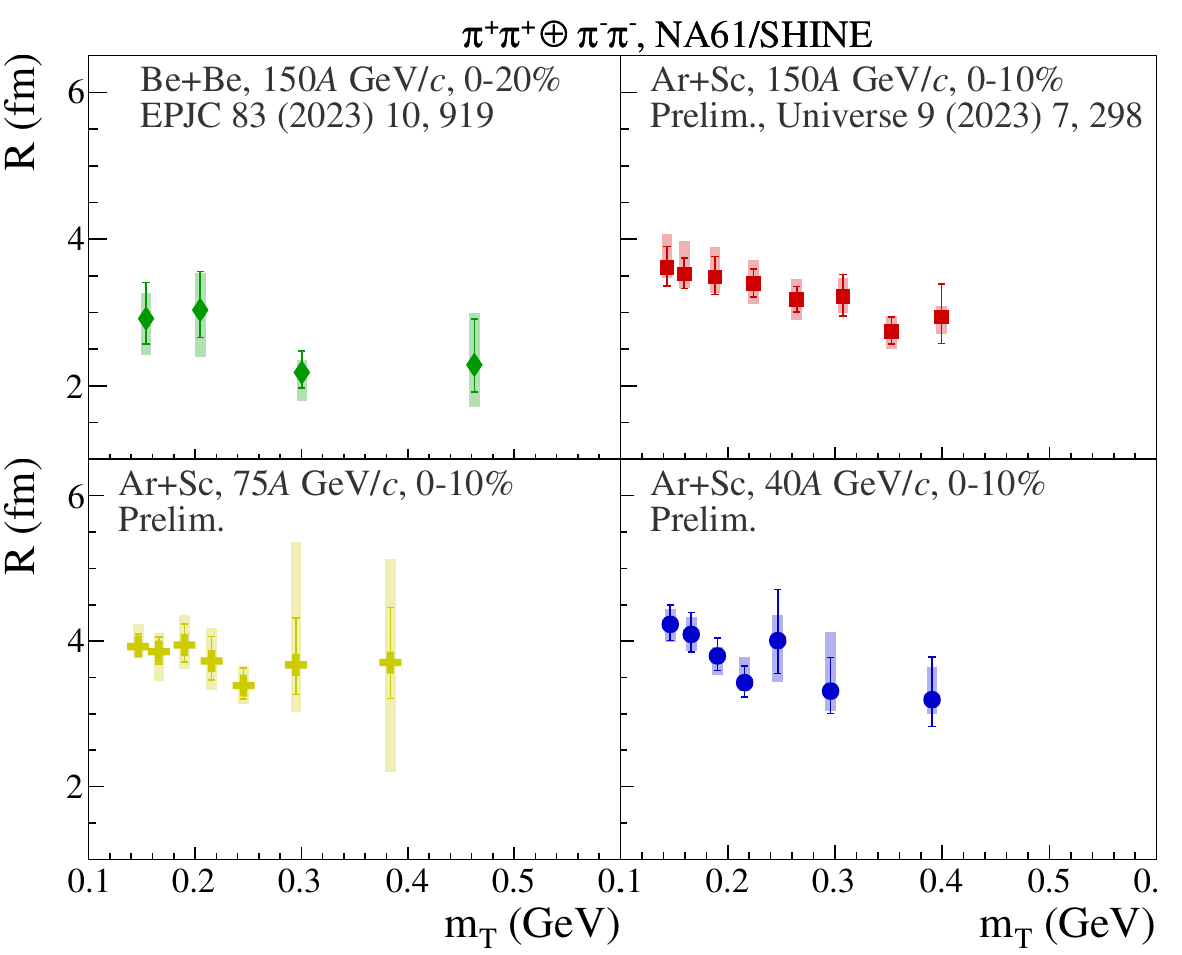}}
\caption{The fit parameters, for $0$--$20$ \% central Be+Be at 150\textit{A} GeV/\textit{c} and $0$--$10$\% central Ar+Sc at 150\textit{A} GeV/\textit{c}, 75\textit{A} GeV/\textit{c}, and 40\textit{A} GeV/\textit{c}, as a function of $m_{\rm{T}}$. Boxes denote systematic uncertainties, bars represent statistical uncertainties.\label{fig:resultsR}}
\end{figure}

The L\'evy scale parameter $R$ is related to kind of homogeneity scale (usually referred as homogeneity length for Gaussian sources~\cite{Sinyukov1995}) of the pion emitting source. From simple hydrodynamical models ~\cite{Csorgo:1995bi, Csanad:2009wc} one obtains a decreasing trend of Gaussian radii with transverse-mass.  In Fig. ~\ref{fig:resultsR} a slight decrease for higher $m_{\rm{T}}$ values are observed for the L\'evy scale $R$ in both systems and in all energies. This might be the result of collective flow. 
Moreover, we also observe an $R(m_{\rm{T}})$ trend compatible with $R \sim 1/\sqrt{m_{\rm{T}}}$ prediction, which is interesting as this particular type of dependence was calculated for Gaussian sources ($\alpha = 2$) in Ref.~\refcite{Sinyukov:1994vg}. 
This phenomenon was also observed at RHIC~\cite{PHENIX:2017ino,Kincses:2024sin} and LHC~\cite{Csanad:2024wlz} and in simulations (at RHIC and LHC energies)~\cite{Kincses:2022eqq,Korodi:2022ohn}. Furthermore, parameter values seem to be similar at all energies in the Ar+Sc system and slightly smaller in the Be+Be system which might be attributed to the difference in initial geometrical size of the two collision systems. 

\begin{figure}[b]
\centerline{\includegraphics[width=1.\textwidth]{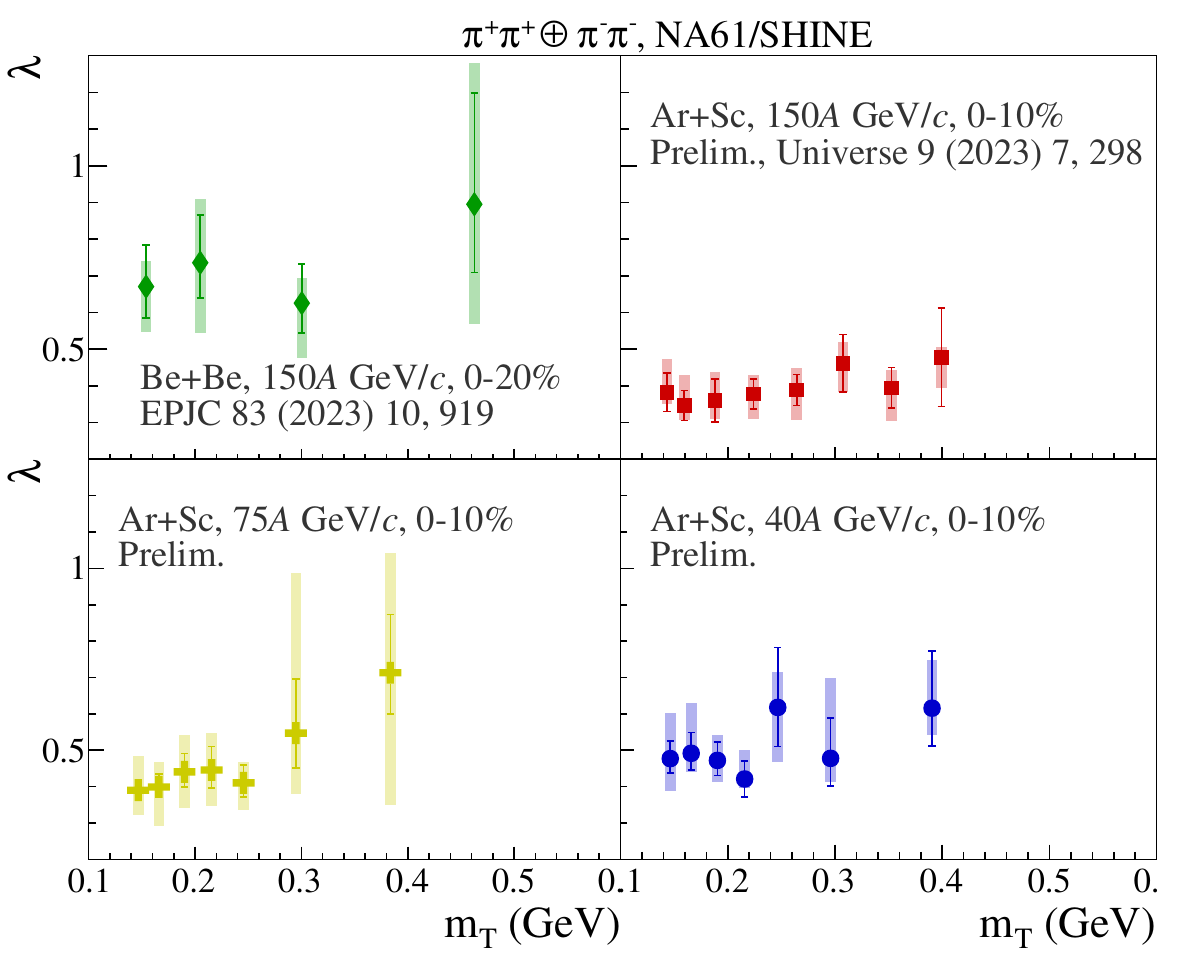}}
\caption{The fit parameters, for $0$--$20$ \% central Be+Be at 150\textit{A} GeV/\textit{c} and $0$--$10$\% central Ar+Sc at 150\textit{A} GeV/\textit{c}, 75\textit{A} GeV/\textit{c}, and 40\textit{A} GeV/\textit{c}, as a function of $m_{\rm{T}}$. Boxes denote systematic uncertainties, bars represent statistical uncertainties.\label{fig:resultsL}}
\end{figure}

The final parameter to investigate is the correlation strength parameter $\lambda$, defined in Eq{.}(\ref{eq:lambda}). The transverse mass dependence of $\lambda$ is shown in Fig. ~\ref{fig:resultsL}. One may observe that this parameter is slightly dependent on $m_{\rm{T}}$ but mostly constant within uncertainties. When compared to measurements from RHIC Au+Au collisions~\cite{PHENIX:2017ino, Vertesi:2009wf, STAR:2009fks} and from SPS Pb+Pb interactions~\cite{Beker:1994qv, NA49:2007fqa}, an interesting phenomenon is observed. In the case of SPS experiments, there is no visible decrease at lower $m_{\rm{T}}$ values, however it appears in the case of RHIC experiments. The decrease (or ``hole'') was interpreted in Ref. ~\refcite{PHENIX:2017ino} to be a sign of in-medium mass modification of $\eta'$. Our results, at the given statistical precision, do not indicate the presence of such a low-$m_{\rm{T}}$ decrease, although in 75\textit{A} GeV/$c$ there seems to be some dependence on $m_{\rm{T}}$, however still within uncertainties. Furthermore, it is important to note that our values for $\lambda$ are significantly smaller than unity, which might imply that a significant fraction of pions are the decay products of long-lived resonances or weak decays.

\section{Conclusion}
In this paper, we presented the NA61/SHINE measurements of one-dimensional two-pion femtoscopic correlation functions in Ar+Sc collisions at 40\textit{A} GeV/\textit{c} and 75\textit{A} GeV/\textit{c} beam momentum within the 
0\--10\% centrality range. We fitted these correlation functons based on the assumption of L\'evy-shaped source distributions, and investigated the transverse mass dependence of the source parameters. We furthermore compared them to previous NA61/SHINE results (150\textit{A} GeV/\textit{c} in both $^{7}$Be+$^{9}$Be and $^{40}$Ar+$^{45}$Sc). Our analysis revealed that the L\'evy scale parameter,  $\alpha$, deviates from Gaussian sources and does not align with the conjectured value at the critical point. Additionally, the L\'evy scale parameter, $R$, exhibits a noticeable decrease with $m_{\rm{T}}$. Unlike RHIC results but consistent with earlier SPS measurements, the correlation strength parameter, $\lambda$, shows no significant dependence on $m_{\rm{T}}$.  In future studies, we plan to measure Bose-Einstein correlations in larger systems and at lower energies to further investigate the phase diagram of strongly interacting matter.


\section*{Acknowledgments}

The author acknowledges support of the DKOP-23 Doctoral Excellence Program of the Ministry for Culture and Innovation, and was furthermore supported by K-138136 and K-138152 grants of the National Research, Development and Innovation Fund.

\bibliography{bporfy_Zimanyi23_LevyHBT_proceedings.bib}
\bibliographystyle{ws-ijmpa.bst}
\end{document}